\documentclass[10pt,nofootinbib,floatfix,a4,showpacs,showkeys,notitlepage]{revtex4-1}

\pdfoutput=1

\usepackage{amsmath}
\usepackage{amssymb}
\usepackage{graphicx}
\usepackage{hyperref}
\usepackage[T1]{fontenc} 
\usepackage{slashed}

\allowdisplaybreaks

\newcommand{\AddrLiege}{%
  IFPA, Dep. AGO, Universit\'e de Li\`ege, Bat B5, Sart-Tilman B-4000
  Li\`ege 1, Belgium}

\begin{document}

\title{Explaining the CMS Higgs flavor violating decay excess}

\author{D. Aristizabal Sierra}
\email{daristizabal@ulg.ac.be}

\author{A. Vicente}
\email{avelino.vicente@ulg.ac.be}

\affiliation{\AddrLiege}
\begin{abstract}
  Direct searches for lepton flavor violating Higgs boson decays in
  the $\tau \mu$ channel have been recently reported by the CMS
  collaboration. The results display a slight excess of signal events
  with a significance of 2.5$\sigma$, which translates into a
  branching ratio of about $1$\%.  By interpreting these findings as a
  hint for beyond the standard model physics, we show that the
  Type-III 2HDM is capable of reproducing such signal while at the
  same time satisfying boundedness from below of the scalar potential,
  perturbativity, electroweak precision data, measured Higgs standard
  decay modes and low-energy lepton flavor violating constraints. We
  have found that the allowed signal strength ranges for the $b\bar
  b$, $WW^*$ and $ZZ^*$ standard channels shrink as soon as
  BR$(h\to\tau\mu)\sim 1$\% is enforced. Thus, we point out that if
  the excess persists, improved measurements of these channels may be
  used to test our Type-III 2HDM scenario.
\end{abstract}
\pacs{12.60.-i, 12.60.Fr}
\keywords{models beyond the standard models, Higgs sector extensions}
\maketitle

\section{Introduction}
Since the discovery of the Higgs boson
\cite{Chatrchyan:2012ufa,Aad:2012tfa} special effort has been made to
determine its properties. The motivation for such an effort resides on
understanding the mechanism for electroweak symmetry breaking.  At
present, several aspects of the Higgs boson are to some extent well
known, in particular those related with some of its expected
``standard'' decay modes, namely: $WW^\ast$, $ZZ^\ast$,
$\gamma\gamma$, $b\bar b$ and $\tau\bar\tau$ . Currently, measurements
of these decay modes have shown compatibility with the standard model
(SM) expectations, although with large associated uncertainties
\cite{CMS:2014ega}.  Indeed, it is due to these large uncertainties
that there is still room for nonstandard decay properties, something
that has encouraged such searches at the LHC as well. Searches for
invisible Higgs decays have been published in
\cite{Aad:2014iia,Chatrchyan:2014tja}, while direct searches for
lepton flavor violating Higgs decays ($h\to\tau\mu$) have been
recently reported by the CMS collaboration in
\cite{lfvHiggsdecays}. In this letter we focus on the latter, for
which the CMS collaboration, using the 2012 dataset taken at
$\sqrt{s}=8$~TeV with an integrated luminosity of 19.7~fb$^{-1}$, has
found a 2.5$\sigma$ excess in the $h\to \tau \mu$ channel, which
translates into BR$(h\to\tau\mu) = \left( 0.89_{-0.37}^{+0.40}
\right)$\%.

Indirect bounds on Higgs lepton flavor violating decay modes arise
from low-energy data. Muon and tau rare decays (e.g. $\mu\to e
\gamma$, $\mu\to 3e$, $\tau\to e\gamma$ and $\tau\to 3\mu$)---induced
by Higgs lepton flavor breaking couplings---place upper bounds on the
Higgs flavor violating modes: $h\to\tau\mu$, $h\to\tau e$, $h\to\mu
e$. Since muon decays have the most tight limits, it is for the $\mu
e$ mode for which consistency with low-energy data demands a branching
fraction well below the LHC reach ($< 10^{-8}$). Constraints on tau
rare processes, being less stringent, allow larger $\tau e$ and
$\tau\mu$ branching ratios \cite{Blankenburg:2012ex,Harnik:2012pb},
hence stimulating these searches at the LHC~\footnote{The impact of
  lepton flavor violating couplings in the Higgs sector on several
  low-energy processes was recently studied in
  \cite{Goudelis:2011un,Celis:2013xja}.}.

Although these bounds follow from a fairly model-independent analysis
(see \cite{Harnik:2012pb} and references therein), one may also wonder
what type of frameworks are capable of producing sizeable lepton
flavor violating Higgs decays. Efforts in such direction have been
done in different contexts, with pioneer works in
Refs. \cite{Pilaftsis:1992st,DiazCruz:1999xe}. More recenty,
Ref. \cite{Arana-Catania:2013xma} studied the problem in the MSSM,
while \cite{Arhrib:2012mg} in the R-parity violating MSSM. Flavor
violating decays have been considered as well in the inverse seesaw
model in \cite{Arganda:2014dta}. Possible effects due to vectorlike
leptons have been investigated in \cite{Falkowski:2013jya}.  Extended
scalar sectors involving several Higgs doublets and flavor symmetries
(Yukawa textures) have been examined too
\cite{Bhattacharyya:2010hp,Bhattacharyya:2012ze,Arroyo:2013kaa,Campos:2014zaa}
\footnote{Minimal flavor violating and Froggatt-Nielsen frameworks
  \cite{D'Ambrosio:2002ex,Froggatt:1978nt} have been investigated
  using an effective approach in Ref. \cite{Dery:2014kxa}. It has been
  pointed out that in their most simple versions either schemes lead
  to nonobservable effects.}. Finally, the Type-III Two Higgs Doublet
Model (2HDM) has been considered in
Refs. \cite{Davidson:2010xv,Kopp:2014rva}. Basically, the bottom line
of all these analyses is that unless one deals with extra Higgs
doublets, lepton flavor violating Higgs decays are below the LHC
reach.

In this paper, we study the viability of producing the CMS excess
signal in the Type-III 2HDM. For simplicity, we assume flavor
violation only in the lepton sector \footnote{For flavor violating
  effects in the quark sector see
  e.g. Refs. \cite{Atwood:1996vj,Crivellin:2013wna}.}  and adopt a
pure phenomenological approach, that is to say all the parameters of
the model are treated as free parameters subject only to
phenomenological constraints. The phenomenological restrictions we
consider are the following. First of all, the new degrees of freedom
are constrained by electroweak precision data. Thus, in our analysis
we calculate the contributions to the $T$ parameter (contributions to
the $S$ and $U$ parameters in the 2HDM are small
\cite{O'Neil:2009nr}). Since the couplings that induce $h\to \tau\mu$
induce as well tau rare processes, we consider the restrictions
arising from $\tau\to\mu\gamma$ (for which we include Barr-Zee
diagrams contributions \cite{Barr:1990vd} as done in
\cite{Davidson:2010xv}), $\tau\to3\mu$ and $\tau\to \eta\mu$. We also
study whether our scenario can explain the $(g-2)_\mu$ discrepancy. In
contrast to previous studies in the 2HDM, we also verify that the
Higgs standard properties deviate only within the allowed measured
ranges \cite{CMS:2014ega}. These include the following Higgs decay
channels: $\tau\bar\tau$, $b\bar b$, $WW^*$ and $ZZ^*$. We check as
well for theoretical constraints, namely that the scalar potential is
bounded from below and contains perturbative parameters.

The rest of the paper is organized as follows. In sec.
\ref{sec:2hdm-description} we discuss basic aspects of the Type-III
2HDM and introduce the formulas employed in our numerical
calculation. In sec. \ref{sec:pheno} we describe the strategy followed
in our numerical analysis and present our results. Finally, in
sec. \ref{sec:conclusions} we summarize and present our conclusions.
\section{Type-III 2HDM}
\label{sec:2hdm-description}
We consider a 2HDM \cite{Lee:1973iz,Branco:1980sz} of
Type-III. Contrary to other versions of the 2HDM, the Type-III 2HDM
does not include any discrete symmetry that serves to distinguish
between Higgs doublets. Therefore, both Higgs doublets are allowed to
couple to all fermion species.

It is common to present the 2HDM in an arbitrary basis in Higgs
space. Instead, following \cite{Davidson:2010xv}, we prefer to
introduce the model in the so-called Higgs basis. In this basis, only
one Higgs doublet acquires a vacuum expectation value (VEV) and the
scalar potential of the model (assuming CP conservation) is given
by~\footnote{We follow the conventions of
  \cite{Davidson:2005cw,Davidson:2010xv} and denote the potential
  parameters in the Higgs basis in upper case.}
\begin{eqnarray}
\mathcal{V} = && M_{11}^2 H_1^\dagger H_1 + M_{22}^2 H_2^\dagger H_2 - \left( M_{12}^2 H_1^\dagger H_2 + \, \text{h.c.} \right) \nonumber \\
&& + \frac{\Lambda_1}{2} \left(H_1^\dagger H_1\right)^2 + \frac{\Lambda_2}{2} \left(H_2^\dagger H_2\right)^2 + \Lambda_3 \left(H_1^\dagger H_1\right)\left(H_2^\dagger H_2\right) + \Lambda_4 \left(H_1^\dagger H_2\right)\left(H_2^\dagger H_1\right) \nonumber \\
&& \left[ \frac{\Lambda_5}{2} \left(H_1^\dagger H_2\right)^2 + \Lambda_6 \left(H_1^\dagger H_1\right)\left(H_1^\dagger H_2\right) + \Lambda_7 \left(H_2^\dagger H_2\right)\left(H_1^\dagger H_2\right) + \, \text{h.c.} \right] \, ,
\end{eqnarray}
where
\begin{equation} \label{h1h2def}
H_1 = \left( \begin{array}{c}
G^+ \\
\frac{1}{\sqrt{2}} \left( v + \varphi_1^0 + i \, G^0 \right)
\end{array} \right) \quad , \quad H_2 = \left( \begin{array}{c}
H^+ \\
\frac{1}{\sqrt{2}} \left( \varphi_2^0 + i \, A \right)
\end{array} \right) \, ,
\end{equation}
are the Higgs doublets in the Higgs basis, such that $\langle H_1^0
\rangle = v/\sqrt{2}$ and $\langle H_2^0 \rangle = 0$. Here
$\varphi_1^0$ and $\varphi_2^0$ are CP-even neutral Higgs fields, $A$
is a CP-odd neutral Higgs field, $H^+$ is a charged Higgs field and
$G^+$ and $G^0$ are Goldstone bosons. Since we assume CP conservation,
$A$ is the physical pseudoscalar Higgs and does not mix with
$\varphi_1^0$ and $\varphi_2^0$. The relation between $\varphi_1^0$
and $\varphi_2^0$ and the scalar mass eigenstates $h$ and $H$ (with
$m_h < m_H$) is
\begin{eqnarray}
h &=& \sin (\beta - \alpha) \, \varphi_1^0 + \cos (\beta - \alpha) \, \varphi_2^0 \label{hHrot1} \, , \\
H &=& \cos (\beta - \alpha) \, \varphi_1^0 - \sin (\beta - \alpha) \, \varphi_2^0 \, . \label{hHrot2}
\end{eqnarray}
Here we have introduced $\beta - \alpha$, the physical mixing angle
that relates the Higgs basis and the mass basis for the CP-even scalar
states. The potential parameters are related to the physical Higgs
masses as
\begin{eqnarray}
m_{H^+}^2 &=& M_{22}^2 + \frac{v^2}{2} \Lambda_3 \, , \label{eqfirst} \\
m_A^2 - m_{H^+}^2 &=& - \frac{v^2}{2} \left( \Lambda_5 - \Lambda_4 \right) \, , \\
m_H^2 + m_h^2 - m_A^2 &=& v^2 \left( \Lambda_1 + \Lambda_5 \right) \, , \\
(m_H^2 - m_h^2)^2 &=& \left[ m_A^2 + \left( \Lambda_5 - \Lambda_1 \right) v^2 \right]^2 + 4 \, \Lambda_6^2 v^4 \, , \\
\sin \left[ 2(\beta - \alpha) \right] &=& - \frac{2 \, \Lambda_6 v^2}{m_H^2 - m_h^2} \, . \label{eqlast}
\end{eqnarray}

We now turn to the Yukawa interactions of the model. In the Higgs
basis for the Higgs doublets and the mass basis for the fermions, the
Yukawa Lagrangian of the model can be written as
\begin{eqnarray}
- \mathcal{L}_Y = && \sqrt{2} \left( \overline{q_L}_j \widetilde H_1 \frac{K_{ij}^\ast m_i^U}{v} {u_R}_i + \overline{q_L}_i H_1 \frac{m_i^D}{v} {d_R}_i + \overline{\ell_L}_i H_1 \frac{m_i^E}{v} {e_R}_i \right) \nonumber \\
&& + \, \overline{q_L}_i \widetilde H_2 \rho_{ij}^U {u_R}_j + \overline{q_L}_i H_2 \rho_{ij}^D {d_R}_j + \overline{\ell_L}_i H_2 \rho_{ij}^E {e_R}_j + \, \text{h.c.} \, .
\end{eqnarray}
Here we denote $\widetilde H_a = i \sigma_2 H_a^\ast$, the fermions
$\left( u_L , d_L , e_L , u_R , d_R , e_R \right)$ are mass
eigenstates, $K$ is the CKM matrix and $i,j = 1,2,3$ are generation
indices. $m_i^U$, $m_i^D$ and $m_i^E$ are the up-type quark, down-type
quark and charged lepton masses, respectively, and $\rho^U$, $\rho^D$
and $\rho^E$ are general $3 \times 3$ complex matrices in flavor
space. For simplicity, in the following we will assume that the $\rho$
matrices are hermitian. Now, using Eqs. \eqref{h1h2def},
\eqref{hHrot1} and \eqref{hHrot2} we can rewrite the leptonic part of
$\mathcal{L}_Y$ as
\begin{eqnarray}
- \mathcal{L}_Y^{\text{leptons}} = && \bar e_i \left( \frac{m_i^E}{v} \delta_{ij} s_{\beta - \alpha} + \frac{1}{\sqrt{2}} \rho_{ij}^E c_{\beta - \alpha} \right) e_j \, h \nonumber \\
&& + \, \bar e_i \left( \frac{m_i^E}{v} \delta_{ij} c_{\beta - \alpha} - \frac{1}{\sqrt{2}} \rho_{ij}^E s_{\beta - \alpha} \right) e_j \, H \nonumber \\
&& + \, \frac{i}{\sqrt{2}} \bar e_i \rho_{ij}^E \gamma_5 e_j \, A + \left[ \bar \nu_i \left( U^\dagger \rho^E \right)_{ij} P_R e_j \, H^+ + \text{h.c.} \right] \, , \label{eqYuk}
\end{eqnarray}
where $U$ is the PMNS matrix, $P_R = (1 + \gamma_5)/2$ is the usual
right-handed chirality projector, $s_{\beta - \alpha} = \sin (\beta -
\alpha)$ and $c_{\beta - \alpha} = \cos (\beta - \alpha)$. From
Eq. \eqref{eqYuk} we can extract the Higgs couplings to fermions
\begin{eqnarray}
g_{hff^\prime} &=& \frac{m_f}{v} s_{\beta - \alpha} \delta_{f f^\prime} + \frac{\rho_{f f^\prime}}{\sqrt{2}} c_{\beta - \alpha} \, , \label{coup1} \\
g_{Hff^\prime} &=& \frac{m_f}{v} c_{\beta - \alpha} \delta_{f f^\prime} - \frac{\rho_{f f^\prime}}{\sqrt{2}} s_{\beta - \alpha} \, , \label{coup2} \\
g_{Aff^\prime} &=& \pm i \gamma_5 \frac{\rho_{f f^\prime}}{\sqrt{2}} \, , \label{coup3}
\end{eqnarray}
where $g_{Aff^\prime} = + \, i \gamma_5 \frac{\rho_{f
    f^\prime}}{\sqrt{2}}$ for down-type quarks and charged leptons and
$g_{Aff^\prime} = - \, i \gamma_5 \frac{\rho_{f f^\prime}}{\sqrt{2}}$
for up-type quarks. Finally, we will consider a specific structure for
the $\rho$ matrices inspired in the \emph{Cheng-Sher ansatz}
\cite{Cheng:1987rs}. First, we normalize $\rho^E$ as
\begin{equation}
\rho_{ij}^E = - \kappa_{ij} \tan \beta_\tau \sqrt{\frac{2 m_i m_j}{v^2}} \, , 
\end{equation}
where $m_i \equiv m_i^E$ and $\beta_\tau$ is the physical mixing angle
defined by the ratio~\footnote{In the Type-III 2HDM there is no unique
  definition of $\tan \beta$ since one can always apply rotations in
  Higgs space, acting also on the Higgs VEVs, $v_1$ and
  $v_2$. Therefore, a specific (physical) definition is required
  \cite{Davidson:2005cw}. Since we are interested in tau flavor
  violation, we choose to define $\tan \beta_\tau$ as the relative
  size of the tau Yukawa coupling and $\sqrt{2} m_\tau/v$, in analogy
  to the usual definition of $\tan \beta$ in the Type-II
  2HDM. \label{foot:tanb}}
\begin{equation}
\tan \beta_\tau = \frac{- \rho_{\tau \tau}}{\sqrt{2} m_\tau/v} \, .
\end{equation}
Note that, by definition, $\kappa_{\tau \tau} = 1$. However, the other
$\kappa_{ij}$'s, and in particular, $\kappa_{\tau \mu} = \kappa_{\mu
  \tau}^\ast$, are free parameters. For the quark $\rho$ matrices we
assume the following Type-II values
\begin{equation} \label{ansatz}
\rho_{ij}^D = - \sqrt{2} \tan \beta_\tau \frac{m_i^D}{v} \delta_{ij} \quad , \quad 
\rho_{ij}^U = \sqrt{2} \cot \beta_\tau \frac{\left( K^\dagger m^U \right)_i}{v} \delta_{ij} \, .
\end{equation}
This parameterization of the $\rho$ matrices is not the most general
one but, as we will see, it leads to results in good agreement with
the experimental constraints. Furthermore, this ansatz can be
understood as a minimal correction beyond the Type-II 2HDM, with the
only departure in the $\tau \mu$ coupling~\cite{Kanemura:2005hr}.

Finally, the Higgs couplings to gauge bosons are fully dictated by the
gauge symmetry. One has $C_{hWW} = s_{\beta - \alpha} \,
C_{hWW}^{\text{SM}}$, $C_{HWW} = c_{\beta - \alpha} \,
C_{hWW}^{\text{SM}}$ and $C_{AWW} = 0$. The couplings to a pair of
$Z$-bosons follow the same proportionality.

\section{Phenomenological analysis}
\label{sec:pheno}
We now proceed to describe our phenomenological analysis. Our results
are based on a random scan of the parameter space, with the following
ranges for the relevant input parameters:
\begin{align}
  \label{eq:parameter-ranges}
  m_H &\subset \left[ 200 , 1000 \right] \, \text{GeV}\ , \qquad
  m_A \subset \left[ 400 , 1000 \right] \, \text{GeV}\ ,
  \\
  m_{H^\pm} &= m_A + \delta m \ , \qquad \text{with}\qquad
  \delta m \subset \left[ -5 , 5 \right] \, \text{GeV}\ , \\
  \sin (\beta - \alpha) &\subset [0.7 , 1.0]\ ,\qquad
  \tan \beta_\tau \subset [0.1 , 40]\ , \\
  |\kappa_{\tau \mu}| &\subset [0.1 , 3]\ . \label{eq:parameter-ranges-last}
\end{align}
In addition, we assume $m_h = 126$ GeV. Since we are driven by
phenomenological considerations, we use as input the scalar masses
($m_h$, $m_H$, $m_A$ and $m_{H^\pm}$) rather than the parameters in
the scalar potential. Our choice of a small mass difference between
the pseudoscalar and charged Higgses is motivated by the reduction of
the $T$ parameter\footnote{An alternative choice is given by
  $m_{H^\pm}=m_{H^0}+\delta m$, see \cite{Gerard:2007kn} for details.}
, whereas the lower limits on $\sin (\beta - \alpha)$ and $m_{H^\pm}$
(as induced by the lower limit on $m_A$) are motivated by experimental
constraints: smaller values of $\sin (\beta - \alpha)$ would be
excluded by current measurements of the Higgs couplings to fermions
and gauge bosons while lower $m_{H^\pm}$ would lead to certain tension
with flavor physics bounds (mainly B physics, see
\cite{Crivellin:2013wna} for a review). We emphasize that the mass
ranges selected for our numerical scan are quite conservative since
lower values for $m_H$, $m_A$ and $m_{H^\pm}$ are allowed by LHC data,
see e.g. \cite{Celis:2013ixa}.

After choosing input parameters three checks are performed:
\begin{itemize}
\item We determine $\Lambda_1$, $\Lambda_4$, $\Lambda_5$ and
  $\Lambda_6$ by means of Eqs. \eqref{eqfirst}-\eqref{eqlast} and only
  allow those parameter points where the resulting values are below $4
  \pi$. The exact value of this perturbativity constraint is somehow
  arbitrary (see for example the discussion in
  \cite{Ferreira:2009jb,Eriksson:2009ws,Broggio:2014mna}). However, we
  have checked that it has no impact on our numerical
  results~\footnote{We point out that a more rigorous check based on
    tree-level unitarity was introduced in
    \cite{Ginzburg:2005dt}.}. Furthermore, we also ensure boundedness
  from below \cite{Ivanov:2006yq,Ivanov:2007de,Ferreira:2009jb} by
  properly choosing the remaining $\Lambda$ parameters.

\item We explicitly compute the $T$ parameter and discard parameter
  points outside the current 1 $\sigma$ region, given by $T \in \left[
    -0.03 , 0.19 \right]$ \cite{Baak:2012kk}.

\item Finally, we check whether the Higgs couplings agree with the
  current CMS measurements \cite{CMS:2014ega}. More precisely, we
  determine the signal strengths for $h \to \tau \bar \tau , b \bar b
  , W W^\ast , Z Z^\ast$, defined as $\mu = \left( \sigma \times
  \text{BR} \right) / \left( \sigma \times \text{BR}
  \right)_{\text{SM}}$, and compare them to the CMS 1 $\sigma$ ranges
  \cite{CMS:2014ega}. We assume that Higgs production is given by
  gluon fusion (neglecting other production mechanisms, a fairly good
  approximation) and compute the signal strengths as

\end{itemize} 
\begin{equation}
  \mu_{XX} = \left( \frac{g_{htt}}{g_{htt}^{\text{SM}}} \right)^2
  \left( \frac{g_{hXX}}{g_{hXX}^{\text{SM}}} \right)^2 \ .
\end{equation}
Here $g_{htt}$ and $g_{hXX}$ are the Higgs boson couplings to a pair
of top quarks and $X \equiv \tau , b , W , Z$, respectively. The
superscript SM indicates standard model values.

These constraints impose severe restrictions on the allowed parameter
space of the model (most parameter points are actually excluded) and
their consideration is required to properly address the Type-III 2HDM
as responsible for the CMS $h \to \tau \mu$ signal excess. After these
checks are passed, the observables are computed using the analytical
expressions of \cite{Davidson:2010xv}. In particular, we compute the
branching ratio for $h \to \tau \mu$ as
\begin{equation}
\text{BR}(h \to \tau \mu) = \frac{m_h}{8 \pi \Gamma_h}
\left( |g_{h \tau \mu}|^2 + |g_{h \mu \tau}|^2 \right) \, ,
\end{equation}
where $\Gamma_h$ is the Higgs boson total decay width~\footnote{Given
  the constraints on the Higgs boson couplings, the computed
  $\Gamma_h$ turns out to be close to the SM value,
  $\Gamma_h^{\text{SM}} \simeq 4.1$ MeV.}. For the analytical
expressions required for the computation of BR$(\tau \to \mu \gamma)$
we refer the reader to Appendix \ref{ap:taumugamma}.

\begin{figure}[t!]
  \begin{center} 
    \includegraphics[scale=0.7]{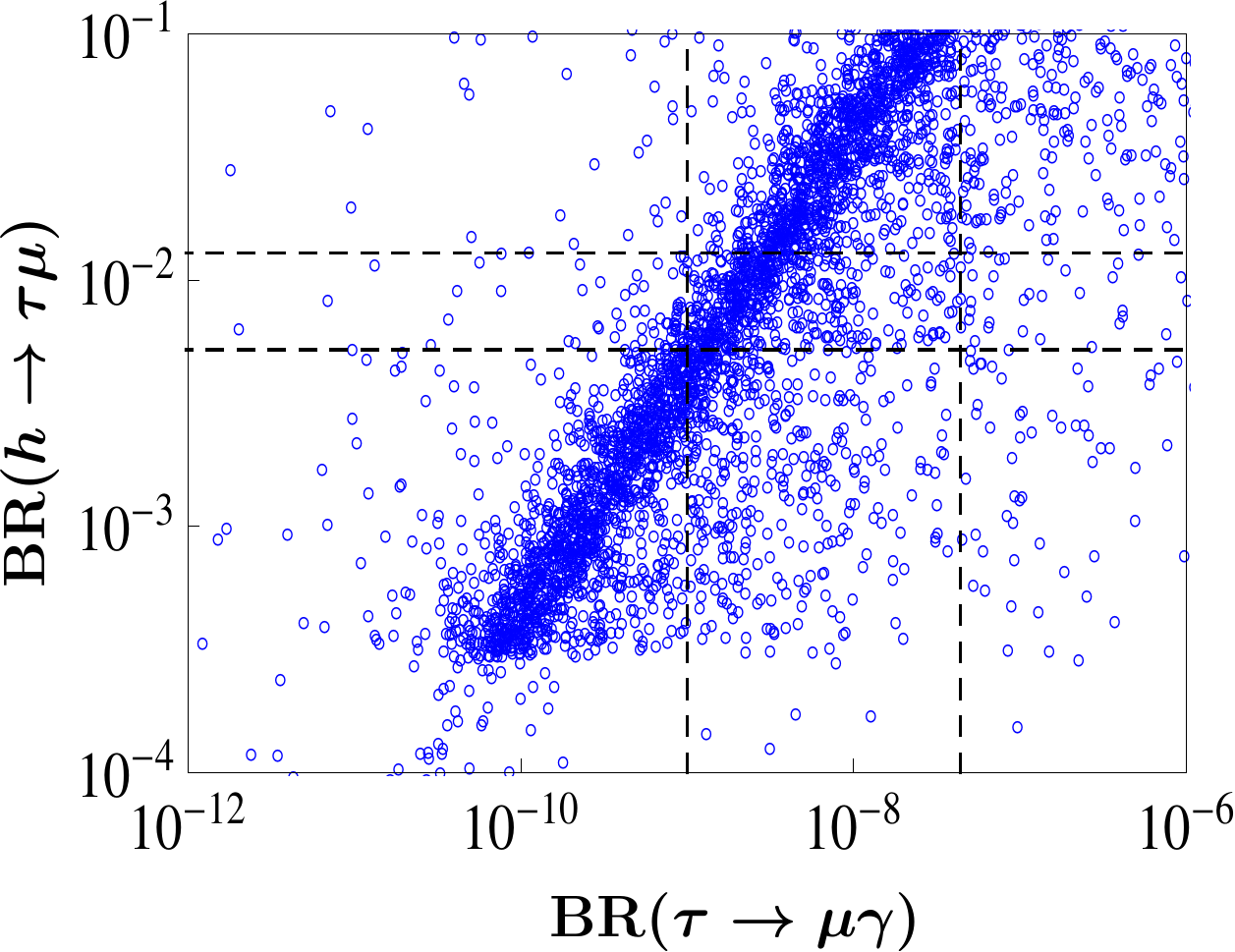}
  \end{center}
  \caption{\small BR$(h \to \tau \mu)$ as a function of BR$(\tau \to
    \mu \gamma)$ for the relevant parameters fixed according to
    Eqs. \eqref{eq:parameter-ranges}-\eqref{eq:parameter-ranges-last}. The
    horizontal lines show the 1 $\sigma$ interval around the best-fit
    value for BR$(h \to \tau \mu)$, as observed by
    CMS~\cite{lfvHiggsdecays}. The right vertical line corresponds to
    the current experimental limit BR$(\tau \to \mu \gamma) < 4.4
    \times 10^{-8}$~\cite{Aubert:2009ag}, whereas the left vertical
    line represents an expected sensitivity at Belle-II of about $\sim
    10^{-9}$~\cite{Aushev:2010bq}.  \label{fig:htmu}}
\end{figure}

Fig. \ref{fig:htmu} shows our results for BR$(h \to \tau \mu)$ as a
function of BR$(\tau \to \mu \gamma)$. The horizontal lines define the
1 $\sigma$ interval around the observed BR$(h \to \tau \mu)$ at
CMS~\cite{lfvHiggsdecays}. The vertical lines show the current
experimental limit set by BaBar, BR$(\tau \to \mu \gamma) < 4.4 \times
10^{-8}$~\cite{Aubert:2009ag}, and an expected sensitivity at Belle-II
of about $\sim 10^{-9}$~\cite{Aushev:2010bq}. As can be clearly seen,
there is a correlation between BR$(h \to \tau \mu)$ and BR$(\tau \to
\mu \gamma)$. However, cancellations among the different contributions
to BR$(\tau \to \mu \gamma)$ preclude any definitive prediction for
this observable. We find that the dominant contribution is typically
given by 2-loop Barr-Zee diagrams \cite{Bjorken:1977br,Barr:1990vd}
with internal $W$ bosons, although the other contributions considered
in our numerical evaluation~\footnote{Following \cite{Chang:1993kw},
  these are 1-loop diagrams with a Higgs boson in the loop and 2-loop
  Barr-Zee diagrams with internal t- and b-quarks, see Appendix
  \ref{ap:taumugamma}.}  may have similar sizes (and even become
dominant in some cases). We conclude that it is possible to explain
the CMS excess in $h \to \tau \mu$, while being compatible with the
current bound on BR$(\tau \to \mu \gamma)$ as well as the
abovementioned constraints on the scalar potential, the Higgs
couplings and the $T$ parameter. This is the main result of this
paper.

Regarding the required values for the model parameters, a sizable
$\kappa_{\tau \mu} \sim 0.5 - 0.8$ is necessary. This is the key
parameter in the determination of BR$(h \to \tau \mu)$. In contrast,
we find little dependence with the other parameters (when taken
individually), although specific ranges are favored by the
experimental constraints. In what concerns $\tan \beta_\tau$, a value
$\tan \beta_\tau \gtrsim 2$ is necessary to be compatible with the
constraints on the Higgs boson couplings. The $\beta - \alpha$ angle
should departure from $\pi/2$ (which would correspond to the
decoupling limit) and have values in the $\sin (\beta - \alpha) \sim
0.9$ ballpark. Finally, we find a preference for large $\Lambda$
parameters, required to generate a hierarchy in the scalar spectrum.

\begin{figure}[t!]
  \begin{center} 
    \includegraphics[scale=0.6]{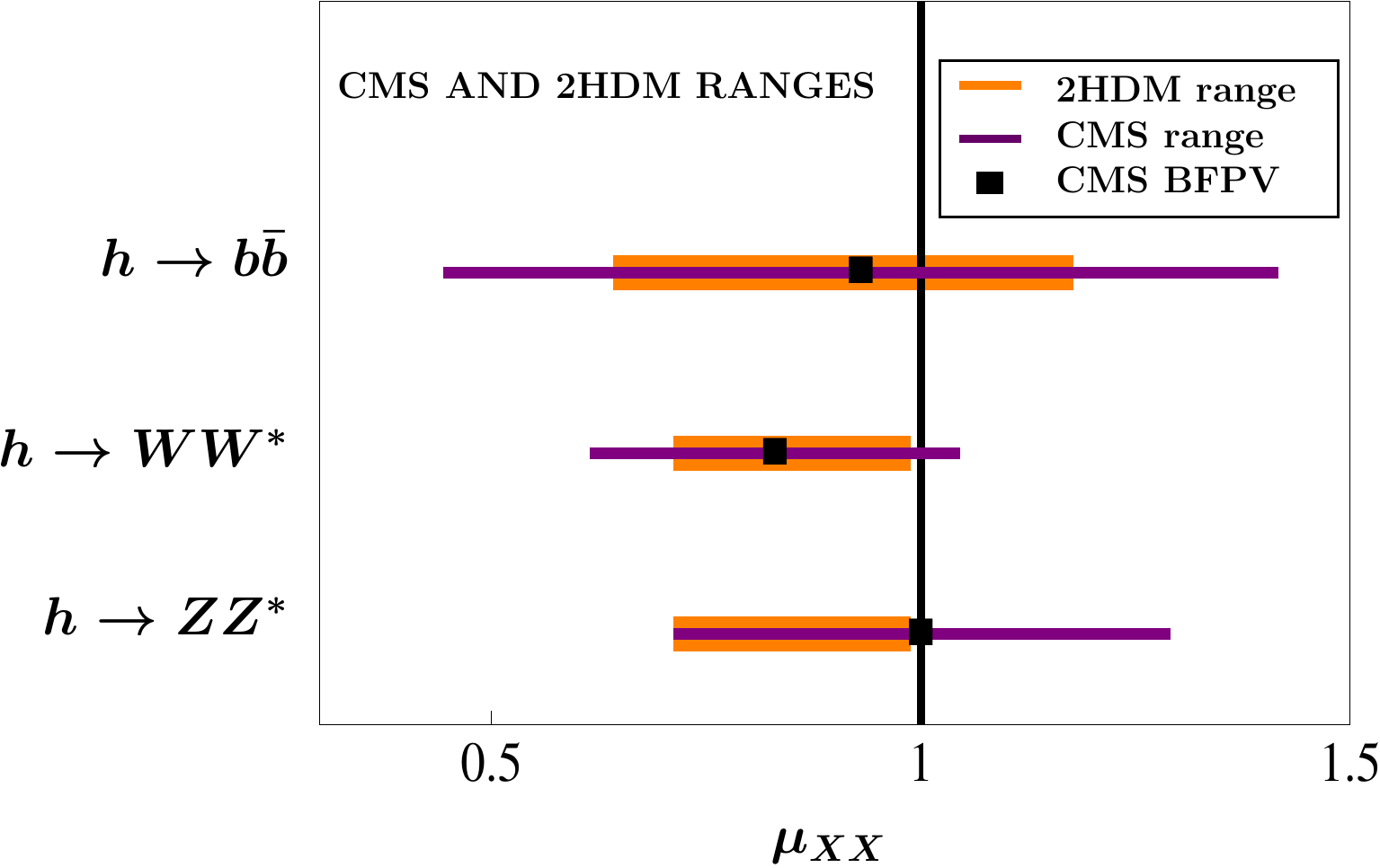}
  \end{center}
  \caption{\small Allowed signal strengths in the Type-III 2HDM as
    required to produce BR$(h\to \tau\mu)$ in agreement with the
    signal found at CMS (within 1 $\sigma$) while at the time
    satisfying electroweak precision data ($T$ parameter constraints)
    as well as low-energy data. The purple (thin) bars cover the 1
    $\sigma$ uncertainties obtained by CMS (with the black dots at the
    best-fit values), whereas the orange (thick) bars represent the
    allowed signal strengths in the Type-III 2HDM as required to
    obtain BR$(h\to \tau\mu) = \left( 0.89_{-0.37}^{+0.40}
    \right)$\%. \label{fig:signal}}
\end{figure}

In Fig. \ref{fig:signal} we present our results for the allowed signal
strengths in the $b \bar{b}$, $W W^\ast$ and $Z Z^\ast$
channels~\footnote{A recent global fit of the Higgs couplings in the
  Type-I and Type-II 2HDM was presented in \cite{Bernon:2014vta}.}. As
usual, they are normalized to their SM values and thus the vertical
line at $\mu_{XX} = 1$ represents the SM prediction. The purple bars
cover the complete 1 $\sigma$ ranges compatible with measurements by
CMS \cite{CMS:2014ega}, with the black dots at the best-fit
values. The orange bars represent the allowed signal strengths in the
Type-III 2HDM as required to obtain BR$(h\to \tau\mu) = \left(
0.89_{-0.37}^{+0.40} \right)$\%. We do not show the results for the
$\tau \tau$ channel because they do not provide any information, as
the purple and orange bars extend over the same range. The most
interesting results are those for the $W W^\ast$ and $Z Z^\ast$
channels, $\mu_{WW},\mu_{ZZ} \in \left[ 0.71 , 0.99 \right]$. For
both, a measurement implying $\mu > 1$ would rule out our scenario. In
contrast, for the $b \bar{b}$ channel both $\mu_{bb} < 1$ and
$\mu_{bb} > 1$ are compatible with the CMS measurement of BR$(h\to
\tau\mu)$, although the allowed bar shrinks to a narrower range,
$\mu_{bb} \in \left[ 0.64 , 1.18 \right]$. Nevertheless, we note that
the results for this channel are related to the ansatz assumed in
Eq. \eqref{ansatz}.

Finally, we have also considered the LFV processes $\tau \to \eta \mu$
and $\tau \to 3 \, \mu$. However, we have found that the constraints
on the model derived from their experimental bounds are weaker than
those obtained from the radiative $\tau \to \mu \gamma$. Furthermore,
our numerical results show that the $(g-2)_\mu$ anomaly cannot be
explained within our scenario, in agreement with
\cite{Davidson:2010xv}.

\section{Conclusions}
\label{sec:conclusions}
The excess of signal events in the $\tau\mu$ Higgs boson decay channel
reported by the CMS collaboration poses a challenge for beyond SM
physics models. Basically, apart from models involving extra Higgs
doublets, no models capable of producing Higgs lepton flavor violating
measurable at LHC have been put forward. Certainly, the 2HDM is the
simplest of such extensions, and in its Type-III ``incarnation''
combines the elements that---in principle---can yield an explanation
to CMS data.

Driven by this motivation, in this paper, we have done a detailed
study of the Type-III 2HDM. In our analysis, we have payed special
attention to all relevant constraints, which include: (i) boundedness
from below and perturbativity of the scalar potential; (ii)
electroweak precision data; (iii) low-energy $\tau$ lepton flavor
violating decay constraints; (iv) standard Higgs decay channels. Our
findings show that generating a large $h\to\tau\mu$ branching
fraction, as required by CMS data, turns out to be possible while
satisfying criteria (i)-(iv). Conditions in (i) do not lead to
significant contraints due to large parameter freedom. Conditions in
(ii) and (iii) do not represent either any unavoidable restriction. We
have found that the most stringent constraints arise from requiring
the Higgs boson to obey current limits on its standard decay channels:
$b\bar b$, $\tau\bar \tau$,$WW^*$, $ZZ^*$, although never threatening
the production of the CMS signal.

In summary, in this paper we have demonstrated that the Type-III 2HDM
with a {\it Cheng-Sher ansatz}-inspired flavor structure naturally
accounts for the CMS excess signal.  Of course, if this excess persist
pinning down its origin will require an extensive experimental
effort. However, we stress that it might be the first indication of a
more complex scalar structure. Moreover, since {\it Cheng-Sher} flavor
structures emerge from flavor symmetries, these data might even reveal
the presence of a fundamental flavor symmetry~\cite{Antaramian:1992ya}.
\section*{Acknowledgements}
DAS and AV want to thank Martin Hirsch for drawing their attention to
the CMS results and Sacha Davidson for the illuminating and inspiring
conversations that led to this work, for her suggestions,
encouragement and comments on the manuscript.  AV is grateful to
Alejandro Celis for fruitful discussions on the subject of Higgs
lepton flavor violation in the 2HDM. We would like to thank as well
Ben O'Leary and Igor Ivanov for comments. DAS would like to
acknowledge financial support from the Belgian FNRS agency through a
``Charg\'e de Recherche'' contract. AV acknowledges partial support
from the EXPL/FIS-NUC/0460/2013 project financed by the Portuguese
FCT.

\appendix

\section{Analytical expressions for $\tau \to \mu \gamma$}
\label{ap:taumugamma}

In this appendix we give the required analytical expressions for the
computation of the $\tau \to \mu \gamma$ rate. This radiative process
is induced by the dipole operator
\begin{equation}
C^{ij} \overline{e_i} \sigma^{\alpha \beta} P_R e_j F_{\alpha \beta} + \text{h.c.} \, ,
\end{equation}
where $i,j$ are the flavors of the external leptons, $F_{\alpha
  \beta}$ is the electromagnetic strength tensor and $\sigma^{\alpha
  \beta} = \frac{i}{2} \left[ \gamma^\alpha , \gamma^\beta
  \right]$. The coefficients $C^{ij}$ can be related to the form
factors $A_L$ and $A_R$,
\begin{equation}
  C^{\tau \mu} = \frac{e \, m_\tau A^{\tau \mu}_{R}}{2} 
  \quad , \quad 
  C^{\mu \tau \ast} = \frac{e \, m_\tau A^{\tau \mu }_{L}}{2} \, ,
\end{equation}
which appear in the $\tau \to \mu \gamma$ branching ratio as
\begin{equation}
\text{BR}(\tau \to \mu \gamma) = \text{BR}(\tau \to \mu \nu \bar{\nu}) \, 
\frac{48 \pi^3 \alpha}{G_F^2} \left( |A_L|^2 + |A_R|^2 \right) \, .
\end{equation}
In the model under consideration $|g_{h \tau \mu}| = |g_{h \mu
  \tau}|$, which leads to $|A_L| = |A_R| \equiv |A|$. As in
\cite{Davidson:2010xv}, we will consider three contributions to the
form factor $A$: 1-loop diagrams with neutral Higgs bosons and charged
leptons in the loop, 2-loop Barr-Zee diagrams with an internal photon
and a third generation quark, and 2-loop Barr-Zee diagrams with an
internal photon and a $W$-boson \cite{Chang:1993kw}. Therefore, we
write
\begin{equation}
A = \frac{1 }{16 \pi^2} \left( A_1 + A_2^{t,b} + A_2^W \right) \, .
\end{equation}
The different contributions are \cite{Davidson:2010xv}
\begin{eqnarray}
A_1 &=& \sqrt{2} \sum_\phi   
 \frac{ g_{\phi \mu \tau}  g_{\phi \tau \tau}}{m_{\phi}^2}
\left( \ln \frac{m_{\phi}^2}{m_\tau^2} - \frac{3}{2} \right) \, , \label{eq:A1} \\
A_2^{t,b} &=& 2  \sum_{\phi,f} g_{\phi \mu \tau} g_{\phi ff}  
\frac{N_c Q_f^2 \alpha}{ \pi}
\frac{1 }{m_\tau m_{f}} \, f_\phi(\frac{m_f^2}{m_{\phi}^2}) \, , \label{eq:A2a} \\
A_2^W &=& - \sum_{\phi = h,H} g_{\phi \mu \tau}  C_{\phi WW}
 \frac{ g \alpha}{2\pi m_\tau  m_W}
\left[ 3 f_\phi(\frac{m_W^2}{m_\phi^2})  + \frac{23}{4} 
g (\frac{m_W^2}{m_\phi^2}) + \frac{3}{4} 
h (\frac{m_W^2}{m_\phi^2})  + m_\phi^2 \frac{f_\phi(\frac{m_W^2}{m_\phi^2})
- g(\frac{m_W^2}{m_\phi^2})}{2m_W^2} \right] \, . \label{eq:A2b}
\end{eqnarray}
Here $\phi = h, H, A$, $f$ = $t,b$, the coupling $g_{\phi ff^\prime}$
of the internal loop fermion to the $\phi$ scalar is given in
Eqs. \eqref{coup1}-\eqref{coup3} and the couplings $C_{\phi WW}$ are
given in the last paragraph of Sec. \ref{sec:2hdm-description}. In
Eqs. \eqref{eq:A1}-\eqref{eq:A2b} lepton masses have been neglected
whenever possible. Finally, the loop functions in the previous
expressions are \cite{Chang:1993kw}
\begin{eqnarray}
  f_A(z) \equiv g(z) &=& \frac{z}{2} \int_0^1 dx  \frac{1}{ x(1-x)-z } \ln \frac{x(1-x)}{z} \, , \\
  f_{h,H}(z) &=& \frac{z}{2} \int_0^1 \, dx \, 
  \frac{ (1-2x(1-x))}{x(1-x)-z} \ln\frac{x(1-x)}{z} \, , \\
  h(z) &=& - \frac{z}{2} \int_0^1 \frac{dx}{x(1-x)-z} \left[1 - \frac{z}{x(1-x)-z}
    \ln\frac{x(1-x)}{z} \right] \, .
\end{eqnarray}

\end{document}